\documentclass{jpp}
\usepackage{graphicx}

\usepackage[utf8]{inputenc}
\usepackage[T1]{fontenc}
\usepackage{amsmath}

\usepackage{xcolor}
\usepackage{url} 
\usepackage{lineno}
\usepackage{soul}
\usepackage{amsmath,bm}

\shorttitle{Whistlers driven by electron trapping in magnetic dips}
\shortauthor{I. Svenningsson, I. Pusztai, K. Steinvall and T. Fülöp}

\title{Whistler instability driven by electron trapping in magnetic dips}

\author{Ida Svenningsson\aff{1}
  \corresp{\email{ida.svenningsson@chalmers.se}},
  István Pusztai\aff{1},
  Konrad Steinvall\aff{1}
 \and Tünde Fülöp\aff{1}}

\affiliation{\aff{1}Department of Physics and Astronomy, Chalmers University of Technology,
Gothenburg, 412 96, Sweden}

\begin{document}

\maketitle

\begin{abstract}
Using fully kinetic particle-in-cell simulations, we show how free energy
released by an ion-scale instability reaches electron scales through its effect
on the electron distribution. 
An
initial ion temperature anisotropy drives partly compressional ion cyclotron
waves, which form magnetic dips. Electrons trapped in these dips develop an enhanced temperature anisotropy and butterfly-shaped distributions, similar to those observed by spacecraft in the Earth's magnetosheath. 
By tracking the energy transfer in velocity space, we show that the gradients created by trapping drive whistler waves through cyclotron resonance.  The whistlers in turn scatter the electrons in pitch angle, flattening velocity-space gradients and heating the electrons along the magnetic field. As a result, the electron anisotropy drops to about half of the double-adiabatic prediction once the whistler power peaks. 
Our results show that it is the shape of the electron distribution in magnetic
dips, rather than its temperature anisotropy alone, that drives the whistlers.

\end{abstract}

\section{Introduction}
\label{sec:intro}
An outstanding question in plasma physics is how energy is transported between
scales. In space plasmas, free energy stored at large scales, for instance in
pressure anisotropies, is often released through ion-scale instabilities. The
resulting magnetic structures can, in turn, drive electron-scale waves. When the
magnetic field strength varies, as in mirror modes or compressional ion cyclotron
waves, electrons get trapped in local field minima (which we will refer to as
magnetic dips), and their electron velocity distribution functions (eVDFs)
become strongly non-Maxwellian \citep{Verscharen2026}. Whistler waves are regularly found inside such dips, from the magnetosheath
``lion roars'' first reported by \citet{SmithTsurutani1976} to recent Magnetospheric Multiscale (MMS) \citep{burch_magnetospheric_2016}
observations \citep{Svenningsson2024}. These
right-handed, circularly polarized waves below the electron cyclotron frequency
play an important role in many plasma environments. They scatter electrons into
the loss cone in the Earth's radiation belts \citep{summers_timescales_2007}, broaden the
anti-sunward strahl beam in the solar wind \citep{cattell_parker_2021}, and limit the electron
temperature anisotropy and heat flux in the solar wind and the Earth's
magnetosheath \citep{coburn_regulation_2024, Svenningsson2024, svenningsson_heatflux_2026}.

Such limits are usually described by instability thresholds derived for
bi-Maxwellian distributions \citep{GaryWang1996}. Inside magnetic dips, however,
the electrons often have butterfly-shaped pitch-angle distributions, with fewer
electrons at field-aligned and perpendicular pitch angles than at oblique angles
\citep{Yao2018, Svenningsson2022}. Such non-Maxwellian features can make the
distribution unstable even when its temperature anisotropy is small, so the
bi-Maxwellian thresholds alone are not enough to predict whistler growth
\citep{Kitamura2020,Svenningsson2022}. A common picture has emerged in which butterfly
distributions form as trapped electrons, conserving their adiabatic invariants,  respond  to an evolving dip.
As the dip deepens, betatron cooling depletes the electrons near $90^\circ$,
while processes that exchange energy between perpendicular and parallel motion,
such as Fermi acceleration as mirror points converge
\citep{kivelson_mirror_1996} and the mirror force acting on electrons moving
along the field, enhance the distribution at intermediate pitch angles.
\citet{Yao2018} found butterfly distributions in MMS observations of
magnetosheath mirror modes and proposed betatron cooling together with the
variation of pitch angles along the field as their origin. Test-particle
simulations by \citet{Li2021}, starting from a kinetic equilibrium of a magnetic
dip, confirmed that betatron cooling alone cannot produce the butterfly shape,
and that the mirror force is essential. These studies, however, either inferred
the evolution of the dip from theory \citep{kivelson_mirror_1996} and
observations \citep{Yao2018}, or prescribed it \citep{Li2021}. They could
therefore not include the feedback of the electrons on the magnetic field, nor
the waves that the resulting distributions may excite.

The link between these distributions and the whistlers has mostly been studied
using spacecraft observations. MMS measurements have identified the source
regions of whistlers near the field minima of mirror-mode structures and shown
that the local electron distributions there are unstable to whistler waves
\citep{Kitamura2020, Svenningsson2022}. \citet{Jiang2024} went a step further
and combined MMS data from a magnetic hole with a quasilinear model. They found
that the whistlers grow mainly through cyclotron resonance with electrons moving
opposite to the waves, which reduces the electron anisotropy and heats the
electrons along the field. Such studies show which distributions are unstable
and how they relax, but not how they were formed.

Simulating the full process, from ion-scale instability to electron-scale waves, self-consistently is challenging, due to the need to resolve electron scales while allowing the time required for ion modes to develop. \citet{Ley2024} did this with fully
kinetic simulations of the mirror instability, and showed that electrons trapped in nonlinear mirror modes develop loss-cone-like distributions that excite whistlers. Their study focused on high-$\beta$ plasma of galaxy clusters, with the anisotropy continuously driven by a shearing box setup.
\citet{Ahmadi2018} considered magnetosheath conditions, using expanding-box
simulations in which mirror modes evolve into magnetic holes, and found that the
electrons become whistler-unstable in the magnetic minima even though they are
stable on average. Their analysis of the simulation was, however, based on the
temperature anisotropy rather than on the full distribution function.
 What remains unclear is how the full shape of the eVDF evolves as
compressive structures grow, how the
whistlers act back on it in velocity space, and how much of the anisotropy
reduction is due to the whistlers rather than to the changing magnetic field.

In this paper, we study this process with fully kinetic 2D3V particle-in-cell simulations, in which the ion-scale structures form self-consistently from an initial ion temperature anisotropy. The anisotropy drives ion cyclotron waves, which grow to large amplitudes and become partly compressional. Electrons are trapped in the magnetic minima of these waves, where they develop butterfly-like distributions. We find that the velocity-space gradients created by the trapping drive whistler
waves, even where the temperature anisotropy is modest, and that the whistlers
propagate away from the minima and scatter electrons in pitch angle.
By comparing with the double-adiabatic prediction, we show that this scattering, and not only the relaxation of the magnetic structures, is what reduces the electron anisotropy. Finally, we compare the electron signatures in the simulation with observations from the MMS mission.

\section{Simulation setup and system evolution}

We simulate the self-consistent development of ion- and electron-scale fluctuations using the fully kinetic particle-in-cell code Smilei \citep{Derouillat2018}, with two spatial dimensions $(x,y)$ and three velocity dimensions. For each species $\alpha\in{e,i}$, we define the inertial length $d_\alpha=c/\omega_{p\alpha}$, where $\omega_{p\alpha}=\sqrt{n_\alpha q_\alpha^2/(\epsilon_0 m_\alpha)}$ is the plasma frequency, and the cyclotron frequency $\Omega_{c\alpha}=|q_\alpha|B_0/m_\alpha$. Here, $n_\alpha$, $q_\alpha$, and $m_\alpha$ denote the species number density, charge, and mass, respectively. The simulation domain spans $L_x\times L_y= 280\,d_e\times280\,d_e$ with periodic boundary conditions, resolved with $1536\times1536$ grid cells and $1000$ particles per cell and species. To make it feasible to resolve electron dynamics over many ion cyclotron periods, we use a reduced mass ratio $m_i/m_e=100$, giving $d_i=10 d_e$ for a quasineutral plasma with singly charged ions. We set $\Omega_{ce}/\omega_{pe}=1/4$.

The plasma is initialized in a uniform magnetic field
$\boldsymbol{B} = B_0 \hat{\boldsymbol{x}}$. We define the parallel
plasma beta as $\beta_{\alpha\parallel}
= 2\mu_0 n_\alpha T_{\alpha\parallel}/B^2$,
where 
$T_{\alpha \parallel}$ is the parallel temperature of species $\alpha$ (in units of energy). The parallel and perpendicular directions
are defined relative to the magnetic field.
The ions follow a bi-Maxwellian distribution 
\begin{equation}
f_{\alpha 0}(v_\parallel,v_\perp)
= n_\alpha
\left(\frac{m_\alpha}{2\pi T_{\alpha\parallel}}\right)^{1/2}
\frac{m_\alpha}{2\pi T_{\alpha\perp}}
\exp\!\left[
-\frac{m_\alpha v_\parallel^2}{2T_{\alpha\parallel}}
-\frac{m_\alpha v_\perp^2}{2T_{\alpha\perp}}
\right],
\end{equation}
with
$\beta_{i\parallel} = 2$ and $T_{i\perp}/T_{i\parallel} = 10$,
while the electrons are isotropic with $\beta_{e\parallel} = 1$ and
$T_{e\perp}/T_{e\parallel} = 1$. We define the electron thermal speed as
$v_{{\rm th},e} = \sqrt{2T_{e\parallel}/m_e}$. With the parameters above,
$v_{{\rm th},e} = \sqrt{\beta_{e\parallel}}\,(\Omega_{ce}/\omega_{pe})\,c = 0.25c$
initially.
The strong ion temperature anisotropy is the
only source of free energy in the system and is chosen to speed up the process of growing the initial ion scale waves, thereby saving computational costs. As a result, any electron anisotropy
and electron-scale waves that appear later must develop self-consistently from
the ion-scale dynamics. Because the magnetic field changes direction during the
simulation, we calculate the parallel and perpendicular temperatures as the appropriate projections of the full pressure tensor with respect to
the local magnetic field, following
\citet{Swisdak2016}.

In the following, we first give an overview of the system evolution
(Section~\ref{sec:evolution}). We then examine how the electron temperature
anisotropy is modulated by the magnetic structures (Section~\ref{sec:anisotropy})
and how trapping shapes the electron velocity distribution
(Section~\ref{sec:evdf}). Next, we characterize the ion- and electron-scale waves
(Section~\ref{sec:waves}), identify quasilinear diffusion signatures of the
electron–whistler interaction (Section~\ref{sec:diffusion}), and show how the
whistlers regulate the electron anisotropy (Section~\ref{sec:whistler_aniso}).

\subsection{System evolution} 
\label{sec:evolution}
\begin{figure}
    \centering
    \includegraphics[width=.8\linewidth]{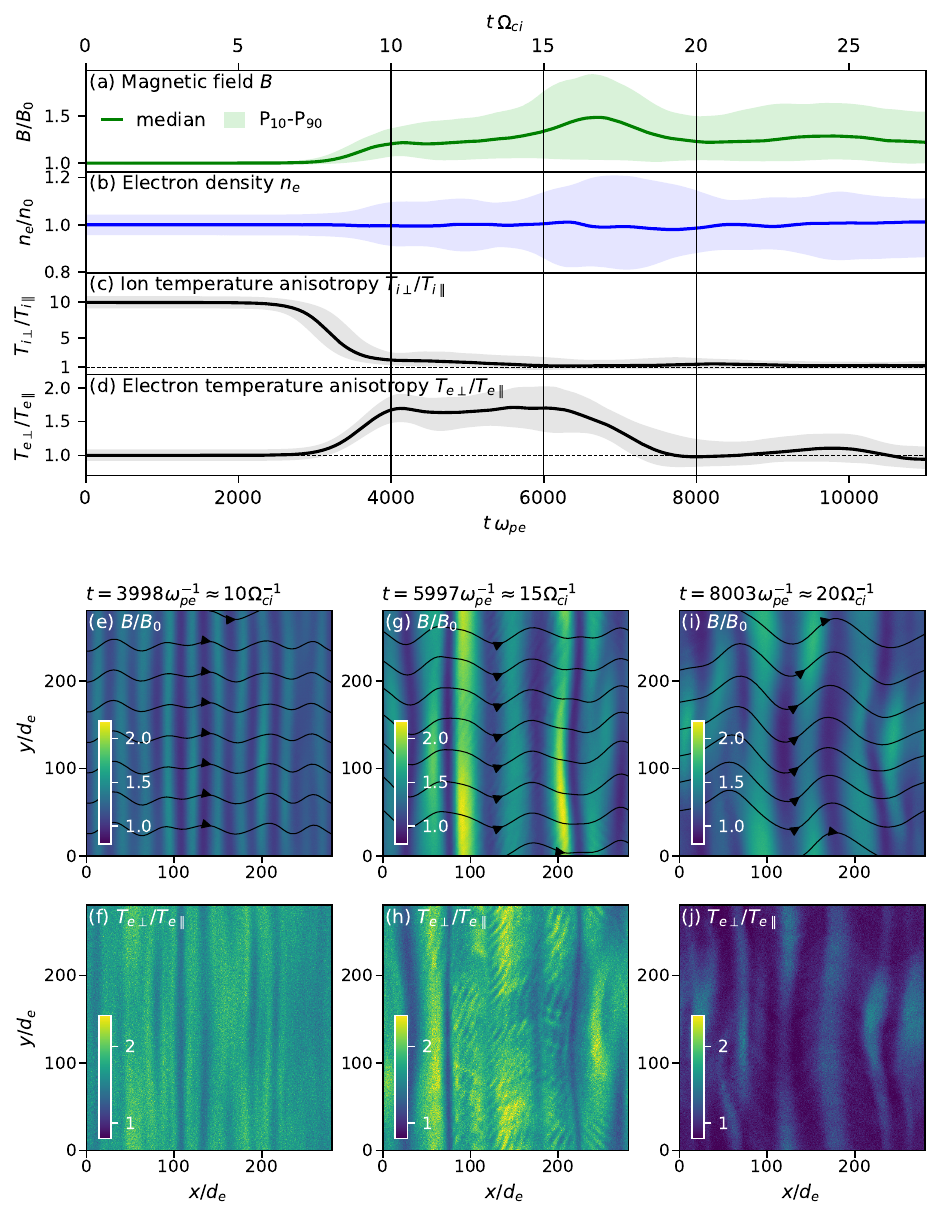}
    \caption{
    Simulation overview. Panels (a)--(d) show the time evolution of (a) the normalized magnetic field strength $B/B_0$, (b) the electron density $n_e$, (c) the ion temperature anisotropy $T_{i\perp}/T_{i\parallel}$, and (d) the electron temperature anisotropy $T_{e\perp}/T_{e\parallel}$. At each time, the solid curves show the spatial medians, and the shaded bands span the 10th to 90th percentiles across the simulation box. The vertical lines mark the times of the snapshots in panels (e)--(j), with $t\omega_{pe} \approx4000$, $6000$, and $8000$ from left to right. The upper row of snapshots (e, g, i) shows $B/B_0$, with streamlines indicating the direction of the magnetic field projected onto the $xy$ plane. The lower row (f, h, j) shows $T_{e\perp}/T_{e\parallel}$.
    }
    \label{fig:system_evolution}
\end{figure}
Figure~\ref{fig:system_evolution}(a)--(d) shows the time evolution of the magnetic field strength, electron density, and ion and electron temperature anisotropies. At each time, the solid curves give the spatial medians, and the shaded bands span the 10th to 90th percentiles across the simulation box. Unlike externally driven shearing- or expanding-box simulations, the system receives no energy input after initialization. The initial magnetic field is uniform and the electric field vanishes, so the energy in the electromagnetic fluctuations that subsequently develop comes from the initial particle kinetic energy provided in the form of the ion temperature anisotropy. Since $T_{i\perp}/T_{i\parallel}$ is the source of energy for the ion-scale
magnetic fluctuations, it decreases as they grow;  the median
$T_{i\perp}/T_{i\parallel}$ falls from 10 to about 2 by $t\omega_{pe} \approx 4000$,
and then approaches unity more gradually (Figure~\ref{fig:system_evolution}c). During the initial drop, the median magnetic field strength rises to about $1.2B_0$ (Figure~\ref{fig:system_evolution}a), and the electrons develop a perpendicular temperature anisotropy, with the median $T_{e\perp}/T_{e\parallel}$ reaching about 1.7 (Figure~\ref{fig:system_evolution}d). The electron temperature anisotropy remains near this value until $t\omega_{pe}\approx6000$, providing a source of free energy for electron-scale fluctuations. A second period of magnetic amplification accompanies the start of the subsequent relaxation of the electron anisotropy. The median field strength peaks at about $1.5B_0$ near $t\omega_{pe}=6700$ and falls to about $1.2B_0$ by $t\omega_{pe}=8000$, when the median electron temperature anisotropy has returned to unity. The median magnetic field strength remains above $B_0$ for the rest of the simulation, while the 10th percentile stays close to $B_0$. The median electron density stays close to its initial value throughout, but its spatial spread broadens during both periods of magnetic amplification and is largest near the second maximum of $B$ (Figure~\ref{fig:system_evolution}b), with the 10th--90th
percentiles reaching about $\pm 20\%$ of the initial density. This shows that the
ion cyclotron waves become compressional and modulate the electron density
significantly, with the density enhanced where the field is weak
(Section~\ref{sec:anisotropy}).

Figure~\ref{fig:system_evolution}(e)--(j) shows how the spatial structure of the magnetic field strength $B$ and the electron temperature anisotropy $T_{e\perp}/T_{e\parallel}$ evolves. By $t\omega_{pe} \approx4000$, following the rapid decrease in ion temperature anisotropy, alternating peaks and dips in $B$ form a nearly periodic pattern of bands elongated along $y$. The in-plane field direction varies periodically along $x$ (Figure~\ref{fig:system_evolution}e). In Section~\ref{sec:waves}, we identify these structures as large-amplitude ion cyclotron waves with a compressional component. The electron temperature anisotropy has risen above unity across most of the domain and exhibits a similar banded pattern (Figure~\ref{fig:system_evolution}f). By $t\omega_{pe} \approx6000$, the magnetic structures are larger and less regular (Figure~\ref{fig:system_evolution}g), and the electron temperature anisotropy varies appreciably within the bands (Figure~\ref{fig:system_evolution}h). At $t\omega_{pe} \approx8000$, magnetic fluctuations are still evident (Figure~\ref{fig:system_evolution}i). Although the median electron temperature anisotropy has returned to approximately unity, substantial local departures from isotropy persist (Figure~\ref{fig:system_evolution}j). In the following sections, we examine how magnetic trapping and adiabatic compression shape the electron distribution.

\subsection{Temperature anisotropy evolution}
\label{sec:anisotropy}

\begin{figure}
    \centering
    \includegraphics[width=\linewidth]{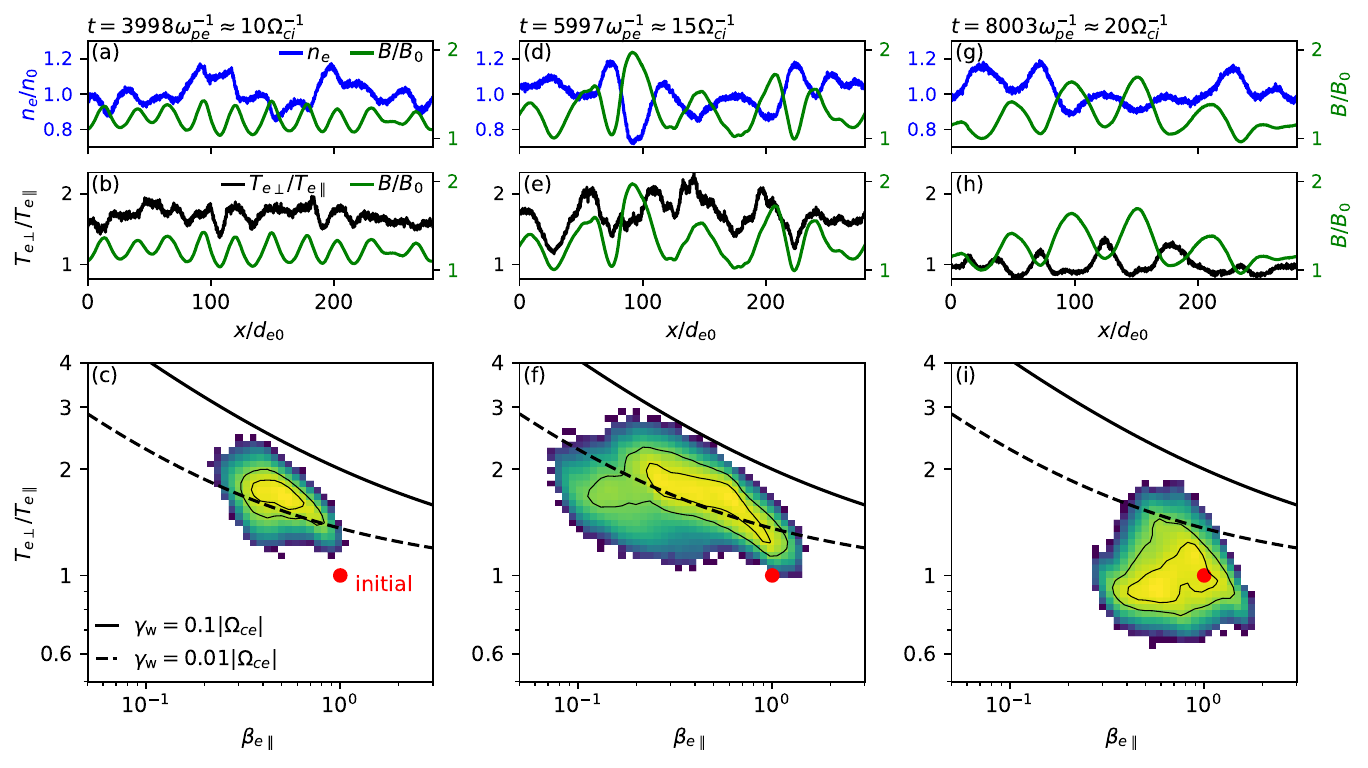}
    \caption{
Electron temperature anisotropy across magnetic structures and its relation to the whistler instability thresholds. From left to right, the columns show $t\omega_{pe} \approx4000$, $6000$, and $8000$. The upper two rows show profiles along $x$, averaged over the strip $y\in(229,231)\,d_e$. The top row (a, d, g) shows the electron density $n_e$ (blue), and the middle row (b, e, h) shows the electron temperature anisotropy $T_{e\perp}/T_{e\parallel}$ (black). Both rows also show the normalized magnetic field strength $B/B_0$ (green, right axis). The bottom row (c, f, i) shows the joint distribution of $\beta_{e\parallel}$ and $T_{e\perp}/T_{e\parallel}$ across the full simulation domain. The thin inner and outer contours enclose 50\% and 90\% of the samples, respectively. The thick solid and dashed curves give the whistler instability thresholds of \citet{GaryWang1996} for growth rates $\gamma_{\rm w}=0.1\,|\Omega_{ce}|$ and $0.01\,|\Omega_{ce}|$, respectively. The red dot marks the initial state $(\beta_{e\parallel},\,T_{e\perp}/T_{e\parallel})=(1,1)$.    
    }
    \label{fig:Brazilplots}
\end{figure}
We now examine how the electron density and temperature anisotropy vary across the magnetic structures. At $t\omega_{pe} \approx4000$, the density fluctuations are weak and have a different
dominant wavelength than $B$, so there is no clear phase relation between them
along the cut in $x$ (Figure~\ref{fig:Brazilplots}a), 
while $T_{e\perp}/T_{e\parallel}$ is generally positively correlated with $B$
along the same cut, although not everywhere (Figure~\ref{fig:Brazilplots}b), consistent with an
adiabatic response to the compression (Section~\ref{sec:whistler_aniso}). 
The data distribution of $\beta_{e\parallel}$ and $T_{e\perp}/T_{e\parallel}$ (Figure~\ref{fig:Brazilplots}c) has shifted toward lower parallel beta and higher electron temperature anisotropy relative to the initial state (red dot). 

At $t\omega_{pe} \approx 6000$, the density fluctuations have grown further and are now
clearly anti-correlated with $B$ (Figure~\ref{fig:Brazilplots}d), with the
density enhanced in the magnetic dips. 
However, the electron temperature anisotropy $T_{e\perp}/T_{e\parallel}$ still has a non-trivial variation with $B$, with local minima at both peaks and dips of $B$. 
The data distribution has broadened, extending from low $\beta_{e\parallel}$ and high $T_{e\perp}/T_{e\parallel}$ toward higher $\beta_{e\parallel}$ and lower $T_{e\perp}/T_{e\parallel}$ (Figure~\ref{fig:Brazilplots}f).  The core of the distribution follows the whistler instability threshold
for $\gamma_w = 0.01\,|\Omega_{ce}|$ \citep[dashed curve;][]{GaryWang1996}.

By $t\omega_{pe} \approx8000$, the earlier positive correlation between $T_{e\perp}/T_{e\parallel}$ and $B$ has given way to a predominantly anticorrelated pattern along the cut, with the largest temperature anisotropies occurring in magnetic dips (Figure~\ref{fig:Brazilplots}h). At the magnetic peaks, $T_{e\perp}/T_{e\parallel}$ falls toward or below unity. The data distribution is concentrated near isotropy, with most samples below the $\gamma_{\rm w}=0.01|\Omega_{ce}|$ contour (Figure~\ref{fig:Brazilplots}i). Thus, even as the bulk of the plasma returns toward isotropy, localized enhancements in $T_{e\perp}/T_{e\parallel}$ persist in magnetic dips.

\subsection{Electron velocity distributions}
\label{sec:evdf}
\begin{figure}
    \centering
    \includegraphics[width=\linewidth]{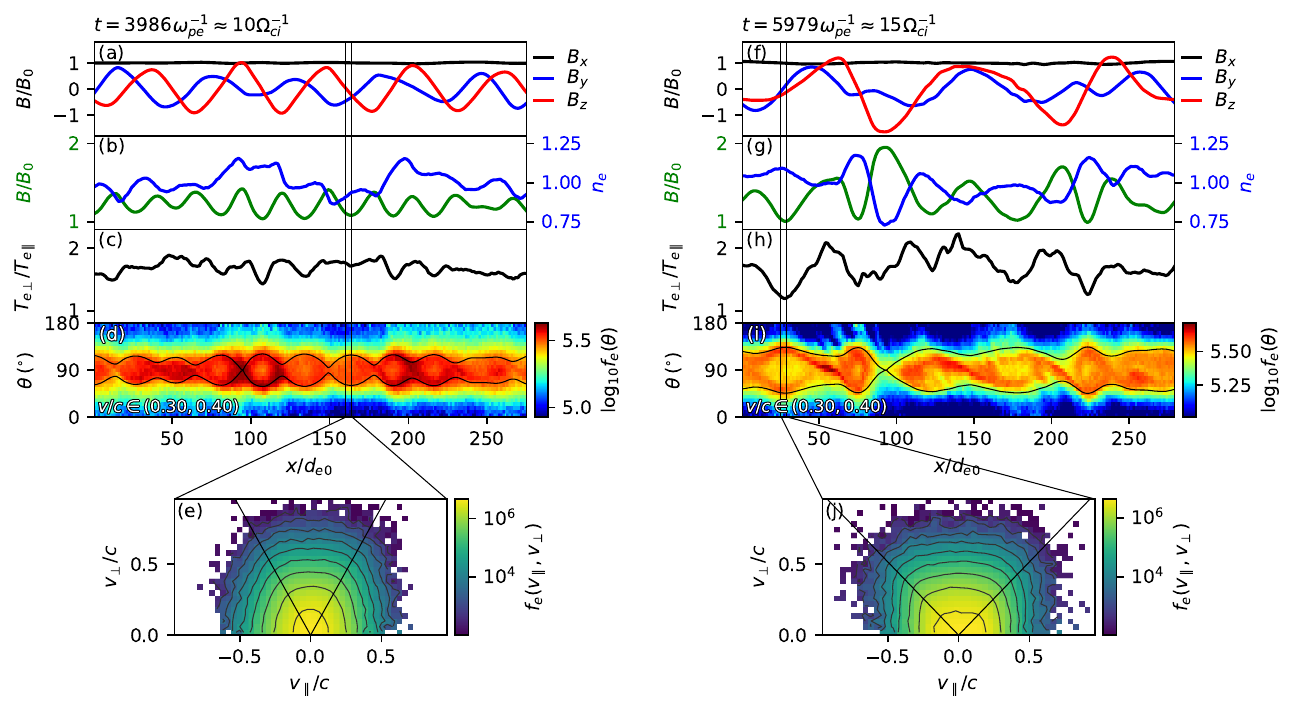}
    \caption{
    Electron trapping in magnetic minima at $t\omega_{pe} \approx4000$ (left column) and $6000$ (right column). The upper four rows show profiles along $x$, averaged over the strip $y\in(228,232)d_e$. Panels (a, f) show the magnetic field components $B_x$ (black), $B_y$ (blue), and $B_z$ (red), normalized to $B_0$. Panels (b, g) show the field strength $B/B_0$ (green, left axis) and electron density $n_e$ (blue, right axis). Panels (c, h) show the electron temperature anisotropy $T_{e\perp}/T_{e\parallel}$. In panels (d, i), color indicates the electron pitch-angle distribution for $v\in(0.3,0.4)c$. The black curves mark the boundaries of the trapped pitch-angle range, $\theta_{\rm tr}$ and $180^\circ-\theta_{\rm tr}$. Panels (e, j) show the electron velocity distribution $f_e(v_\parallel,v_\perp)$ averaged over $x\in(160,164)\,d_e$, $y\in(228,232)\,d_e$ in (e) and $x\in(25,29)\,d_e$, $y\in(228,232)\,d_e$ in (j). The vertical lines in the upper panels mark the corresponding $x$ intervals. In (e, j), the curved black lines are contours of constant $f_e$, and the straight lines mark the trapping boundaries in velocity space.
    }
    \label{fig:eVDF}
\end{figure}

Figure~\ref{fig:eVDF} shows the variation of the electron velocity distribution $f_e$ across
the magnetic structures at $t\omega_{pe} \approx4000$ (left) and $6000$ (right). Figure~\ref{fig:eVDF}a--c shows the magnetic field components, the field strength, the electron density and the temperature anisotropy along $x$ within the strip $y\in(228,232)\,d_e$. Figure~\ref{fig:eVDF}d shows the pitch-angle distribution $f(\theta)$ for $v\in(0.3,0.4)c$. Since $v_{\rm th,e} = 0.25c$, this speed range lies close to the peak of the electron speed distribution, so the pitch-angle structure discussed below affects thermal electrons. It also overlaps with the cyclotron resonance of the whistler waves (see Section~\ref{sec:diffusion}). At $v \approx 0.35c$, the resonant parallel velocities $v_{\parallel,\rm res}/c \in (-0.4, -0.2)$ correspond to pitch angles of about $125^\circ$--$180^\circ$, close to and beyond the trapping boundary.

At $t\omega_{pe} \approx 4000$, there is an overall increase of $f(\theta)$  around $\theta = 90^\circ$, consistent with the high $ T_{e\perp}/T_{e\parallel}$.
We also plot the trapping angle $\theta_{\rm tr}$ (black curves), defined as $\sin^2\theta_{\rm tr}=B/B_{\rm max}$, where $B$ is the local field strength and $B_{\rm max}$ the maximum field strength along $x$. At this time, $f(\theta)$ is not significantly modulated by $\theta_{\rm tr}$ and
the 2D electron velocity distribution $f_e(v_\parallel, v_\perp)$ inside a magnetic dip (Figure~\ref{fig:eVDF}e) is approximately bi-Maxwellian.

Figure~\ref{fig:eVDF}f--j shows a later time step $t\omega_{pe} \approx 6000$ with more clearly anti-correlated magnetic and density fluctuations (Figure~\ref{fig:eVDF}g).
The pitch-angle distribution also varies more, as seen in Figure~\ref{fig:eVDF}i, dropping sharply outside the trapping region $\theta_{\rm tr} < \theta < 180^\circ - \theta_{\rm tr}$.
Apart from the depletion in $f_e(\theta)$ in the parallel and anti-parallel directions, some dips also show a minimum at $\theta = 90^\circ$, which trapping alone does not explain. An example is the dip at $x\in(25,29)\,d_e$ (vertical lines in Figure~\ref{fig:eVDF}f-i; where the full distribution $f_e(v_\parallel, v_\perp)$ is shown in Figure~\ref{fig:eVDF}j).  
Electrons near $90^\circ$ have $v_\parallel \approx 0$ and are not in cyclotron
resonance with the whistlers, so this minimum is not caused directly by whistler
scattering. This type of electron distribution, with local minima at
$\theta = 0^\circ$, $90^\circ$ and $180^\circ$, is called a ``butterfly'' or
``donut'' distribution.

This agrees with the picture outlined in Section~\ref{sec:intro}, in which
butterfly distributions form as trapped electrons respond to an
evolving dip \citep{kivelson_mirror_1996, Yao2018, Li2021}. In our simulation,
the dips indeed deepen between $t\omega_{pe} \approx 4000$ and $6000$. The dip at
$x \in (25, 29)\,d_e$ also shows why bi-Maxwellian instability thresholds are not
sufficient here: despite its pronounced butterfly shape (Figure~\ref{fig:eVDF}j), its
temperature anisotropy is relatively low, about 1.2 (Figure~\ref{fig:eVDF}h). The depletion near
$90^\circ$ reduces the perpendicular pressure, while the depletion near $0^\circ$
and $180^\circ$ reduces the parallel pressure, so the two partly cancel in
$T_{e\perp}/T_{e\parallel}$.

\subsection{Wave characterization}\label{sec:waves}

\begin{figure}
    \centering
    \includegraphics[width=0.8\linewidth]{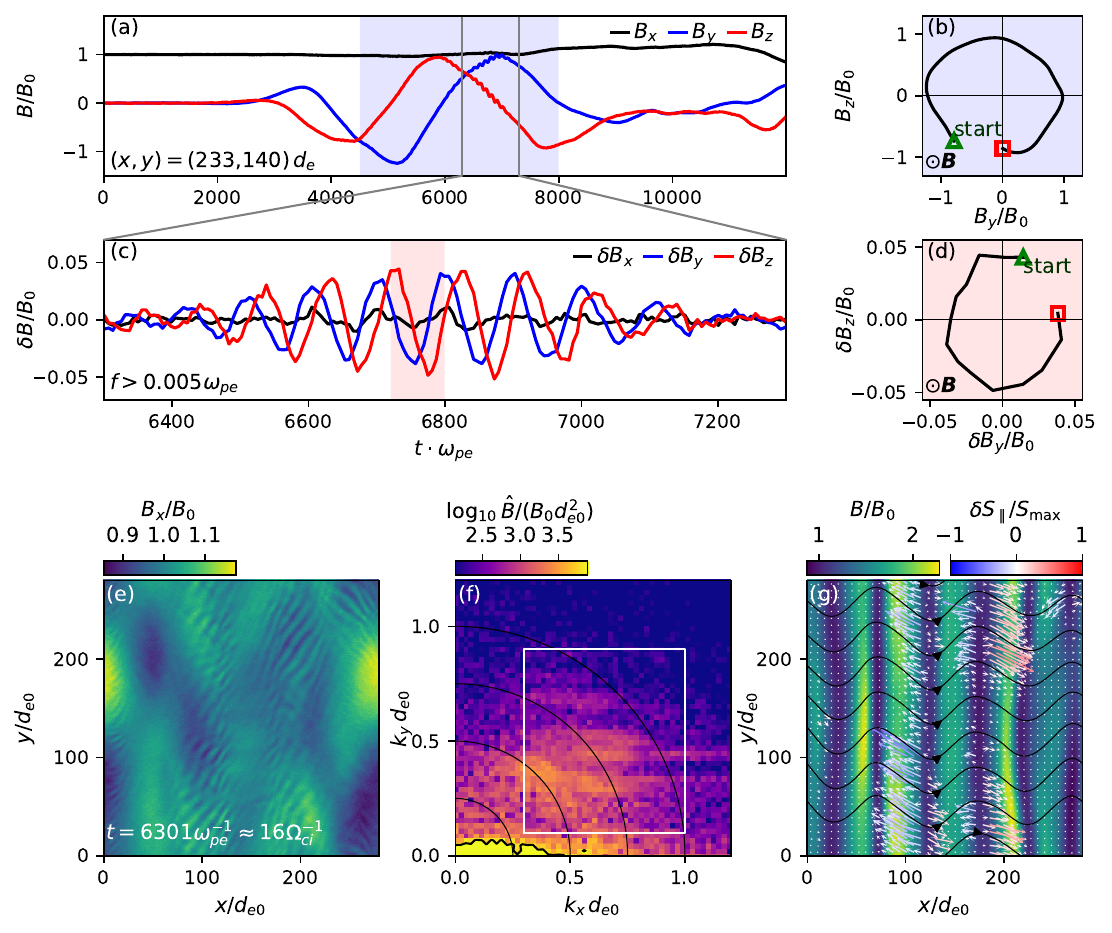}
    \caption{Polarization and propagation of the ion cyclotron and whistler waves.
    (a) Time evolution of the magnetic field components at $x=233\,d_e$, $y=140\,d_e$.
    (b) Hodogram of the $B_y$ and $B_z$ components during the interval marked blue in panel (a), showing the clockwise (left-handed) rotation around the positive $x$ axis.
    (c) High-frequency components $\delta B_{x,y,z}$ ($f>0.005\omega_{pe}$, i.e., $f>0.02|\Omega_{ce}|$) of the field in panel (a) during $t\omega_{pe}\in(6300,7300)$.
    (d) Hodogram of the $\delta B_y$ and $\delta B_z$ components during the interval marked red in panel (c), showing the counter-clockwise (right-handed) rotation of $\delta\bm{B}$ around the positive $x$ axis.
    (e) $B_x/B_0$ at $t\omega_{pe}=6301$.
    (f) Magnitude of the Fourier transform of the magnetic field components, $\hat B = \sqrt{\hat B_x^2 + \hat B_y^2 + \hat B_z^2}$, at the same time. The white box marks the wavenumber range used for filtering the Poynting flux in (g).
    (g) Magnetic field strength $B/B_0$ with arrows showing the Poynting flux $\delta\bm{S}$. The color of the arrows shows the component along the background field, $\delta S_\parallel/S_{\rm max}$, and the black curves show magnetic field lines.
    }
    \label{fig:wave_polarization_propagation}
\end{figure}

To characterize the ion-scale fluctuations, we follow the magnetic field at the point $x=233\,d_e$, $y=140\,d_e$ (Figure~\ref{fig:wave_polarization_propagation}a). The $y$ and $z$ components, perpendicular to the initial magnetic field $\bm{B}$, show quasi-periodic fluctuations which account for the overall increase in $B$ in the system. The hodogram of $B_y$ and $B_z$ in Figure~\ref{fig:wave_polarization_propagation}b shows that the field rotates
in the left-handed sense around $\bm{B}_0$, as expected for ion cyclotron waves.

At higher frequencies, the magnetic field also contains smaller-amplitude
fluctuations, shown in Figure~\ref{fig:wave_polarization_propagation}c. The fluctuations are strongest in the $y$ and $z$ components, as these are close to perpendicular to $\bm{B}$ (see the field lines in Figure~\ref{fig:system_evolution}e). These components have a $90^\circ$ phase shift typical of circular polarization, and the hodogram in
Figure~\ref{fig:wave_polarization_propagation}d shows that they rotate in the
right-handed sense around $\bm{B}_0$. The wave period $T_{\rm w}\approx 100 \, \omega_{pe}^{-1}$ corresponds to an angular wave frequency $\omega_{\rm w}\approx 0.25 \Omega_{ce}$, consistent with whistler waves.

We now focus on the location and propagation of the whistler waves observed within the ion cyclotron wave structures.
In the $B_x$ component (Figure~\ref{fig:wave_polarization_propagation}e), where the ion cyclotron fluctuations have the lowest amplitude, the small-scale whistler wave fronts are directly visible. 
Figure~\ref{fig:wave_polarization_propagation}f shows the 2D Fourier transform of the magnetic field at the same time. The dominating ion cyclotron mode is seen in yellow at the bottom left with $kd_e\sim0-0.5$, corresponding to wavelengths $\lambda> 13\,d_e=1.3\,d_i$. The whistler fluctuations are found at higher wave numbers (shorter wavelengths), dominantly $kd_e\sim0.5-0.8$.

To determine the direction of propagation of the whistlers, we calculate the Poynting flux, $\bm{S}=\bm{E}\times\bm{B}/\mu_0$, which measures the electromagnetic energy flux carried by the fields. For a fluctuating wave field $\delta\bm{E},\delta\bm{B}$, the associated $\delta\bm{S}$ is parallel to the wave group velocity.  We first band-pass filter the electric and magnetic fields at wavenumbers $k_xd_e\in(0.3,1.0), k_yd_e\in(0.1,0.9)$ (white box in Figure~\ref{fig:wave_polarization_propagation}f). In Figure~\ref{fig:wave_polarization_propagation}g, the arrows indicate $\delta\bm{S}$ overplotted on the magnetic field magnitude. The red-white-blue color shows $\delta S_\parallel$, the component of $\delta\bm{S}$ parallel to the background magnetic field direction (here defined as the long wavelength magnetic field component $kd_e<0.2$). The whistlers mainly propagate away from the magnetic minima.
We also note that the wave propagation is approximately parallel to the local magnetic field direction (see the black field lines in Figure~\ref{fig:wave_polarization_propagation}g) even if they are oblique in the simulation box frame 
($k_y>0$; Figure~\ref{fig:wave_polarization_propagation}g).

\begin{figure}
    \centering
    \includegraphics[width=\linewidth]{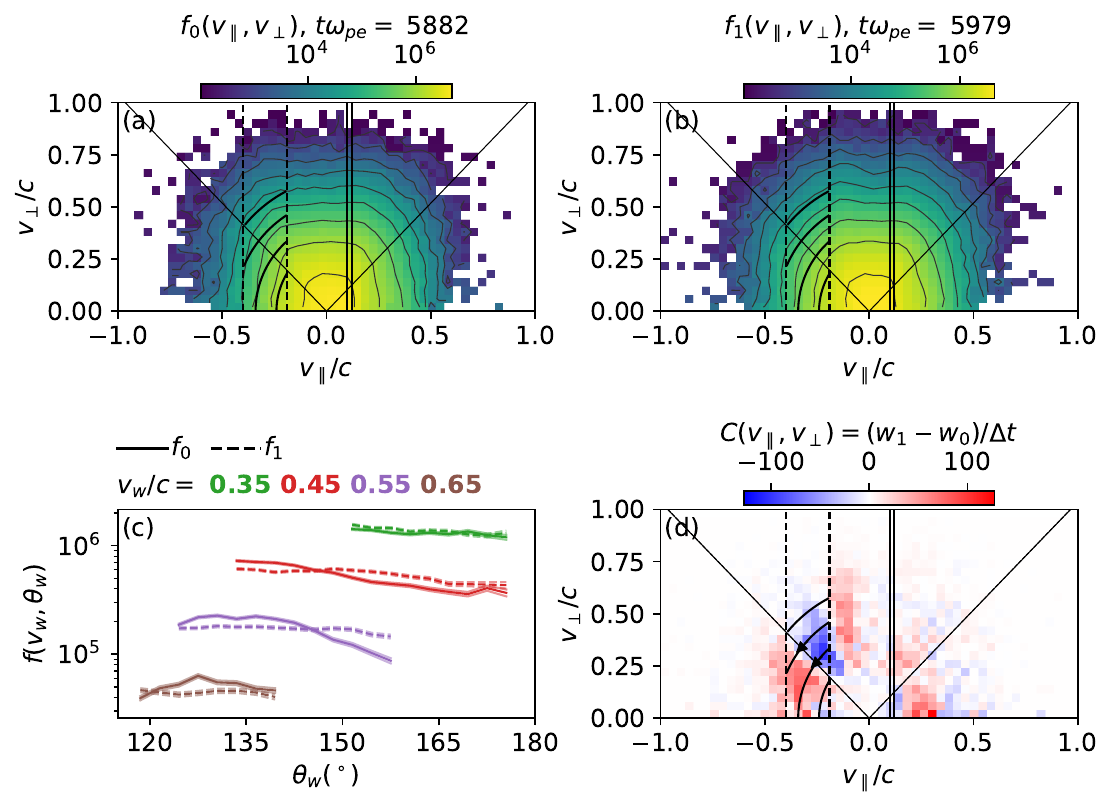}
    \caption{Change in the electron distribution due to whistler waves, averaged
over $x \in (176, 180)\,d_e$, $y \in (228, 232)\,d_e$.
(a) $f_e(v_\parallel, v_\perp)$ at $t\omega_{pe} = 5882$ (denoted here by $f_0$); isocontours are indicated with thin solid black curves. 
     The dashed vertical lines mark the range of resonant velocities
$v_{\parallel\rm res}$, while solid black curves indicate quasilinear diffusion curves within this $v_\|$ range. The solid vertical lines denote the estimated range of the phase velocity
$v_{\parallel\rm ph}$. The thin black diagonal lines show the trapping angle
$\theta_{\rm tr}$.
(b) Same as (a), but at $t\omega_{pe} = 5979$ (denoted here by $f_1$).
(c) $f_0$ (solid) and $f_1$ (dashed) along the quasilinear diffusion curves, where $v_w$ and $\theta_w$ are the speed and pitch angle in the wave frame. The semitransparent curves show the Poisson error $\sqrt{N}$ where $N$ is the number of macro-particles in each bin.
(d) Change in phase-space energy density between (a) and (b),
$C = (w_1 - w_0)/\Delta t$, where $w = m_e v^2 f_e/2$ and
$\Delta t = 97\,\omega_{pe}^{-1}$. Blue and red indicate a decrease and
increase, respectively. The black curves show the diffusion paths of
equation~(\ref{eq:diffusion_curves}) within the resonant range, and the arrows
indicate the direction of diffusion. 
    }
    \label{fig:eVDF_change}
\end{figure}

\subsection{Quasilinear diffusion signatures}
\label{sec:diffusion}
Figure~\ref{fig:eVDF_change} shows how the electron distribution function changes when whistler waves are present. Panels (a) and (b) show $f_e(v_\parallel, v_\perp)$
at $t\omega_{pe} = 5882$ and $5979$, i.e.\ $97\,\omega_{pe}^{-1}$ apart, with
isocontours in black. 
Both distributions are averaged over $x\in(176,180)\,d_e,\, y\in(228,232)\,d_e$. 
This region is at a local magnetic minimum where the whistler waves propagate mainly parallel to $\bm{B}_0$ ($\delta S_\parallel>0$). 

The cyclotron resonance condition defines which regions in $f_e(v_\parallel,v_\perp)$ interact with whistler waves:
\begin{equation}
    v_{\parallel\rm res} = v_{\parallel \rm ph} - l|\Omega_{\rm ce}|/k_\parallel,
\end{equation}
where $v_{\parallel\rm ph} = \omega/k_\parallel$ is the parallel phase
velocity. The $l=1$ first-order cyclotron resonance dominates the energy transfer in the case of quasi-parallel waves \citep{Jiang2024}. We estimate $\omega$ from the cold-plasma dispersion relation,
$k^2c^2/\omega^2 = 1 - \omega_{pe}^2/[\omega(\omega - |\Omega_{ce}|)]$, using
the wavenumber range of the whistlers, $kd_e \in (0.5, 0.8)$
(Figure~\ref{fig:wave_polarization_propagation}f), which coincides with the unstable range found by \citet{Jiang2024} for magnetosheath whistlers.
This gives $v_{\parallel\rm ph}/c\in(0.10,0.12)$ and $v_{\parallel\rm res}^{l=1}/c\in(-0.4,-0.2)$. In Figure~\ref{fig:eVDF_change}, we have indicated these spans with solid and dashed vertical lines, respectively.

We analyze the electron-whistler interaction in the quasilinear diffusion picture, strictly valid for low-amplitude and random-phased wave packets \citep{KennelEngelmann1966,verscharen_electron-driven_2022}. 
 In the frame moving with the phase velocity, the
resonant electrons conserve their energy, so they diffuse along circles centred
at  ${v}_{\rm \parallel ph}$:
\begin{equation}
    (v_\parallel-v_{\parallel \rm ph})^2+v_\perp^2 \equiv v_w^2 = \rm const,
    \label{eq:diffusion_curves}
\end{equation}
shown as black arcs in the $v_{\parallel\rm res}$ range in Figure~\ref{fig:eVDF_change}.  Here $v_w$ is the electron speed in the wave frame,
and we denote the corresponding pitch angle by $\theta_w$.
Figure~\ref{fig:eVDF_change}c shows $f_e$ along these curves for several values of $v_w$. At $t\omega_{pe} = 5882$ (solid), $f_e$ peaks at $\theta_w \approx 135$--$140^\circ$, near the trapping boundary, and falls off toward $\theta_w = 180^\circ$. About $97\,\omega_{pe}^{-1}$ later (dashed), the profiles have become noticeably flatter, as expected when resonant diffusion removes the gradient along the diffusion paths.

By tracking the energy density change in velocity space, we can detect signatures of wave-particle interaction in the electron velocity distribution function \citep{Howes2017, Jiang2024}
\begin{equation}
    C(v_\parallel,v_\perp) = \frac{d}{dt}\left[ \frac{1}{2}m_e v^2f_e(v_\parallel,v_\perp)\right],
\end{equation}
plotted in Figure~\ref{fig:eVDF_change}d. Negative values (blue) correspond to decrease of the phase-space energy density and positive values (red) correspond to increase. Thus, at the boundary between a red and blue region, electrons have diffused from the blue to the red region. 
Within the resonant range, the border between red and blue lies at the
   trapping angle. Electrons leave the trapped population and move along the diffusion curves toward smaller $v_\perp$, which lowers their energy in the simulation frame and releases energy to the waves.
This indicates that the whistlers are driven by
the velocity-space gradient at the trapping boundary, and that the
wave--particle interaction in turn reduces this gradient.

\subsection{Whistler effect on electron anisotropy}
\label{sec:whistler_aniso}
\begin{figure}
    \centering
    \includegraphics[width=\linewidth]{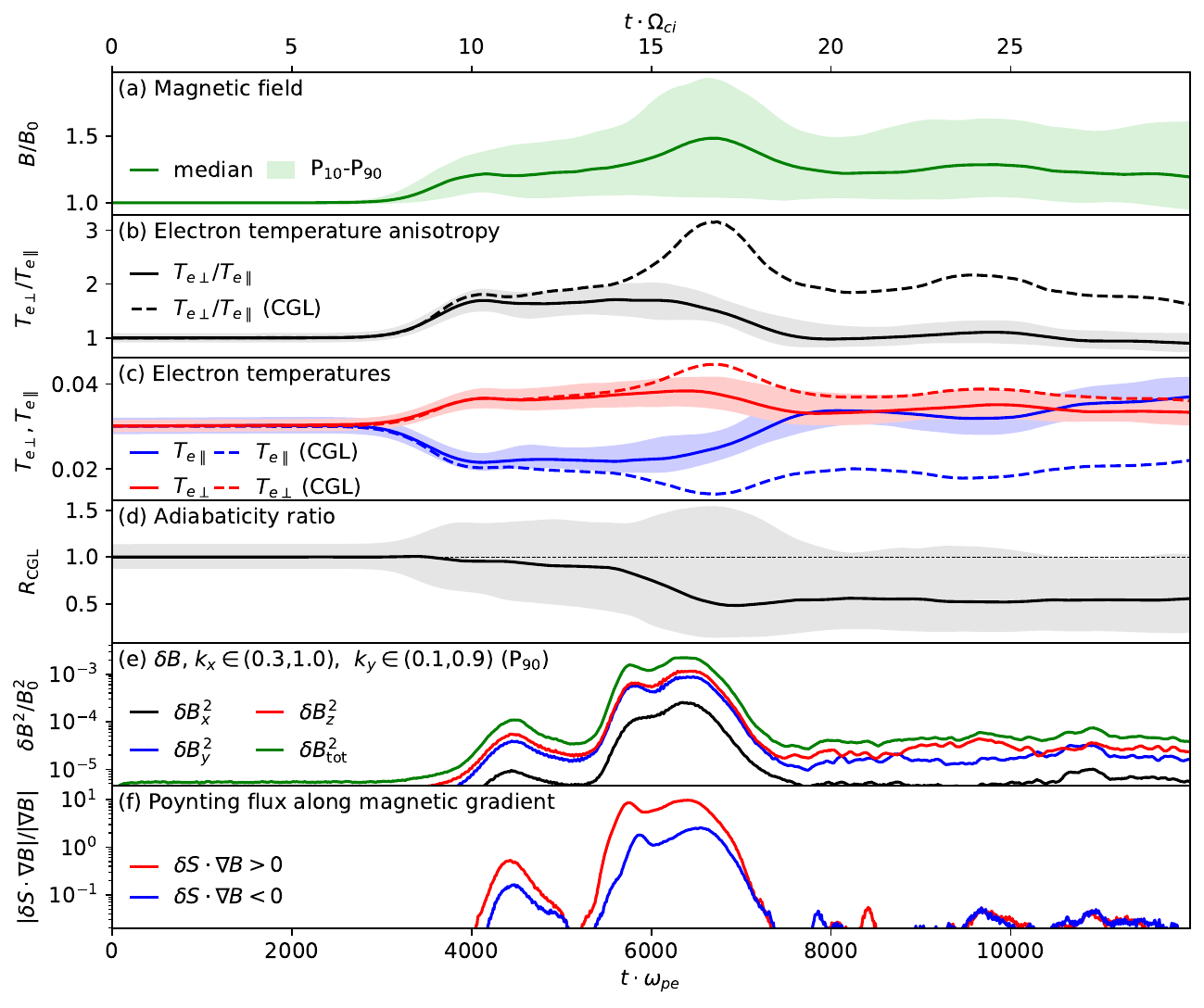}
    \caption{Evolution of the box median (solid) and 10-90th percentiles (shaded) of:
    (a) magnetic field strength $B/B_0$; 
    (b) electron temperature anisotropy ($T_{e\perp}/T_{e\parallel}$);
    (c) electron temperatures parallel ($T_{e\parallel}$) and perpendicular ($T_{e\perp}$) to $\bm{B}$;
    (d) ratio $R_{\rm CGL}$ between the measured and double-adiabatic anisotropy;
    (e) 90th percentile of the magnetic field power in the whistler wavenumber range,
$k_x d_e \in (0.3, 1.0)$, $k_y d_e \in (0.1, 0.9)$; and
 (f) component of the
whistler Poynting flux $\delta\boldsymbol{S}$ along the gradient of the
background magnetic field, $\delta\boldsymbol{S}\cdot\nabla B/|\nabla B|$,
summed over cells where it is positive (red,
toward increasing $B$) and negative (blue, toward decreasing $B$).  Dashed
curves in (b) and (c) show the double-adiabatic (CGL) prediction, equation (\ref{eq:CGL}).}
    \label{fig:timeevolution_waveenergy}
\end{figure}
In Figure~\ref{fig:timeevolution_waveenergy}, we revisit the temporal evolution of the system. 
The whistler activity, measured by the magnetic field power at
$k_x d_e \in (0.3, 1.0)$ and $k_y d_e \in (0.1, 0.9)$
(Figure~\ref{fig:timeevolution_waveenergy}e), develops in two stages. A weaker
enhancement appears at $t\omega_{pe} \approx 4000$, together with the initial
growth of $B$ and $T_{e\perp}/T_{e\parallel}$, and a much stronger burst
follows, peaking at $t\omega_{pe} \approx 6400$ when $B$ is at its maximum and
the trapping is strongest.    A similar two-stage sequence was found by \citet{Ley2024} for mirror modes.

The evolution of magnetic structures changes the parallel and perpendicular electron temperatures even without
any wave--particle interaction. To separate this adiabatic response from the
effect of the whistlers, we compare with the double-adiabatic (Chew–Goldberger–Low, CGL) prediction
\citep{Chew1956},
\begin{equation}
    T_{e\perp}^{\rm CGL} = T_{e0}\,\frac{B}{B_0}, \qquad
    T_{e\parallel}^{\rm CGL} = T_{e0}\left(\frac{n_e}{n_0}\right)^2
    \left(\frac{B_0}{B}\right)^2,
    \label{eq:CGL}
\end{equation}
where $T_{e0}$, $n_0$ and $B_0$ are the initial values. Since the initial state
is uniform, every fluid element starts from the same values, and
(\ref{eq:CGL}) can be evaluated cell by cell from the local $n_e$ and $B$.
The dashed curves in Figure~\ref{fig:timeevolution_waveenergy}b--c show the
resulting box medians, and Figure~\ref{fig:timeevolution_waveenergy}d shows the
ratio between the measured and adiabatic anisotropy,
\begin{equation}
    R_{\rm CGL} = \frac{T_{e\perp}/T_{e\parallel}}
    {\left(T_{e\perp}^{\rm CGL}/T_{e\parallel}^{\rm CGL}\right)}.
    \label{eq:R_CGL}
\end{equation}
Until $t\omega_{pe} \approx 4500$, the electrons follow the adiabatic prediction
closely, and $R_{\rm CGL} \approx 1$. The increase in anisotropy during the
growth of the ion cyclotron waves is therefore mostly an adiabatic response to the
compression of the field, with $T_{e\perp}$ increasing and $T_{e\parallel}$
decreasing. After this, the two diverge. The adiabatic anisotropy keeps growing
and reaches about 3 at $t\omega_{pe} \approx 6700$, whereas the measured
anisotropy stays near 1.7 and then drops to unity by $t\omega_{pe} \approx 8000$.
The median of $R_{\rm CGL}$ falls from about 0.9 to 0.55 between
$t\omega_{pe} \approx 6000$ and $7000$, at the same time as the whistler power
peaks, and remains at this level afterwards. The difference between the adiabatic prediction and the observed values is mainly in the
parallel direction: while the adiabatic prediction gives a roughly constant or
decreasing $T_{e\parallel}$, the measured $T_{e\parallel}$ increases by about
$40\%$ between $t\omega_{pe} \approx 6000$ and $8000$
(Figure~\ref{fig:timeevolution_waveenergy}c). This parallel heating is the
signature expected from pitch-angle scattering by whistlers and agrees with the quasilinear evolution found by \citet{Jiang2024}, as well as our analysis of Figure~\ref{fig:eVDF_change}.

Part of the deviation from the adiabatic prediction may also come from electrons
streaming along the field between dips and peaks. Even if their magnetic moment
is conserved, this redistributes electrons between regions of different field
strength, which the local CGL model, neglecting heat flux, does not include.

Figure~\ref{fig:timeevolution_waveenergy}f shows the
component of the whistler Poynting flux along the gradient of the background
magnetic field, $\delta\boldsymbol{S}\cdot\nabla B/|\nabla B|$, separated into
contributions toward increasing $B$ (red) and decreasing $B$ (blue). The flux
toward increasing $B$ dominates during both whistler bursts: by a factor of about
five during the first burst at $t\omega_{pe} \approx 4300$, by up to an order of
magnitude while the main burst grows, and by a factor of about three at its peak.
The whistlers thus propagate predominantly away from the magnetic dips throughout
the simulation, which supports that the dips are their source regions.

\section{Discussion}
Our simulations show that the whistlers are driven by velocity-space gradients at the
edge of the trapped electron population. Electrons move across the trapping boundary at
the parallel velocities where they resonate with the whistlers, and the
velocity-space gradient is gradually flattened. The Poynting flux points away from the
magnetic minima (Figure~\ref{fig:wave_polarization_propagation}g), so the waves
are emitted from the dips where the trapping takes place.  The comparison with the double-adiabatic prediction shows that the electron
anisotropy follows the adiabatic value while the ion cyclotron waves grow, but
falls well below it when the whistler intensity peaks. 
Without the whistlers, compression
alone would have kept increasing the anisotropy.

\begin{figure}
    \centering
    \includegraphics[width=\linewidth]{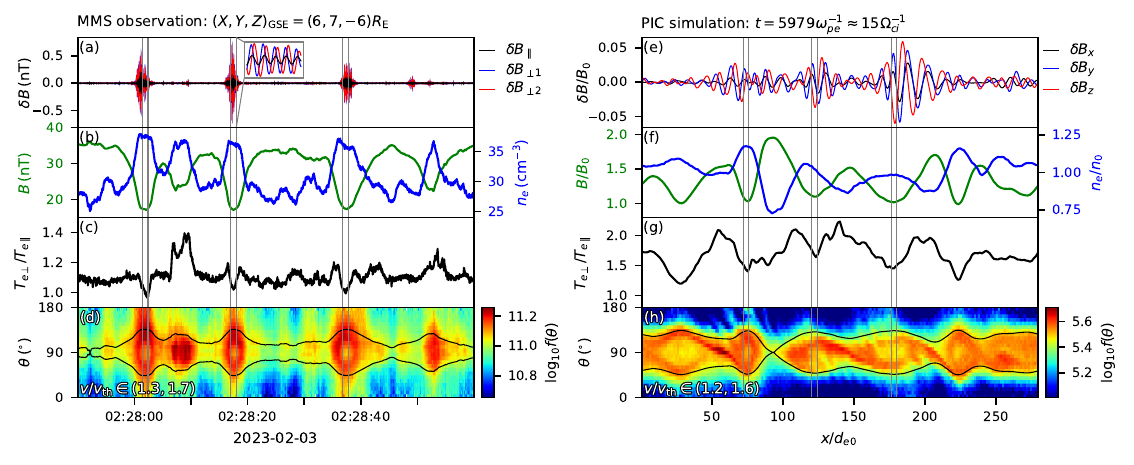}
    \caption{
    (a)-(d) \textit{In situ} observations by the MMS spacecraft \citep{burch_magnetospheric_2016} on 2023-02-03. MMS was in the dayside magnetosheath positioned at $(6,7,-6)R_{\rm E}$, in Geocentric Solar Ecliptic (GSE) coordinates, where $R_{\rm E}\approx 6400\,\rm km$ is the Earth's radius. Time can be read as the spatial coordinate as the structures, except for the high-frequency waves, are approximately stationary in the plasma flowing across the spacecraft.
    (a) High-frequency magnetic field components in the frequency range $(0.05\,|\Omega_{\rm ce}|,|\Omega_{\rm ce}|)$. Whistler bursts, seen mainly in the components perpendicular to the background magnetic field ($\delta B_{\perp1}, \delta B_{\perp2}$), occur in the magnetic dips marked with vertical lines. The inset shows a $50\,\rm ms$ waveform at 02:28:17.
    (b) Magnetic field strength $B$ (green) and electron density $n_e$ (blue), which
are approximately anti-correlated.
    (c) Electron temperature anisotropy $T_{e\perp}/T_{e\parallel}$.
    (d) Electron pitch-angle distribution for $v/v_{{\rm th},e} \in (1.3, 1.7)$, with
the trapping angle $\theta_{\rm tr}$ shown as black curves. 
    (e)-(h) Simulation data at $t\Omega_{ci} = 15$ ($t\omega_{pe} = 5980$) along the
cut in $x$ of Figure~\ref{fig:eVDF}, where the horizontal axis is a spatial
coordinate. 
(e) Small-scale magnetic field components $\delta B_x$, $\delta B_y$ and
$\delta B_z$ in simulation coordinates, normalized to $B_0$ and filtered for wavenumbers $k_xd_e\in(0.3,1.0), k_yd_e\in(0.1,0.9)$.
(f)--(h) Same as Figure~\ref{fig:eVDF}g--i, with the pitch-angle distribution in
(h) shown for $v/v_{{\rm th},e} \in (1.2, 1.6)$.    }
    \label{fig:MMS_vs_PIC}
\end{figure}

The localized diffusion at the trapping boundary indicates a relationship between the magnetic dip and the whistler
frequency. How deep the magnetic dip is, $B/B_{\rm max}$, determines the trapping angle
and through that also the location of the gradient in velocity space. Combined with the resonance condition and the whistler dispersion relation, this determines which
waves can grow. The local electron cyclotron frequency is also lower inside a
dip, which pushes the whistler band to lower frequencies. This is consistent with the MMS observations of \citet{Ahmadi2018}, where the
whistler frequency dropped from $0.35\,|\Omega_{ce}|$ in a shallow magnetic hole
to $0.12\,|\Omega_{ce}|$ in deeper holes.

Although the magnetic dips in our simulation are formed by compressional ion
cyclotron waves rather than mirror modes, the two are closely analogous from the
electrons' perspective. The electron dynamics in a dip depend on the variation of
$B$ along the field lines, through conservation of the magnetic moment and the
mirror force, and less so on the mechanism that produced the dip. Two scale
separations make this picture applicable here. First, the electron gyroradius,
$\rho_e = v_{{\rm th},e}/|\Omega_{ce}| \approx d_e$, is much smaller than the
size of the dips, which extend over tens of $d_e$
(Figures~\ref{fig:Brazilplots} and~\ref{fig:eVDF}), so the magnetic moment is conserved. Second,
the ion cyclotron waves evolve slowly compared with the electron motion. Their
period exceeds $2\pi/\Omega_{ci} \approx 2500\,\omega_{pe}^{-1}$, while a thermal
electron crosses a dip of a few $d_i$ in about $200\,\omega_{pe}^{-1}$. Their phase speed is below the Alfv\'en speed, which is an order of magnitude
smaller than the electron thermal speed,
$v_{{\rm th},e}/v_A = \sqrt{\beta_{e\parallel} m_i/m_e} = 10$. The electrons therefore bounce
many times within a dip while it changes, and experience it as a quasi-static
magnetic bottle, as in a non-propagating mirror mode. With a realistic mass ratio, these scale
separations would be even larger.

The electron signatures in the simulation are similar to those observed by MMS in
the magnetosheath (Figure~\ref{fig:MMS_vs_PIC}). In the MMS data, whistler bursts occur
inside the magnetic dips, mainly in the magnetic field components perpendicular
to the background field (Figure~\ref{fig:MMS_vs_PIC}a). Notably, these bursts occur where the electron pitch-angle distributions exhibit butterfly features: depletion near $0^\circ$, $90^\circ$ and $180^\circ$, and $T_{e\perp}/T_{e\parallel}$ (Figure~\ref{fig:MMS_vs_PIC}c) lower than in the rest of the interval. In the two shallower dips, where $T_{e\perp}/T_{e\parallel}$ peaks, no whistler waves are detected.
This is consistent with our findings in this simulation in that the sharp gradients in trapped electron distributions create the strongest whistler bursts.
In the simulation, the
whistlers are found in and around the dips (Figure~\ref{fig:MMS_vs_PIC}e), consistent
with their propagation away from the dips (Section~\ref{sec:waves}). In both
cases, the density is anti-correlated with the magnetic field strength, the
electron anisotropy varies strongly between dips, and the pitch-angle distributions
inside the dips are depleted near $0^\circ$, $90^\circ$ and $180^\circ$, with the
depletion along the field matching the trapping angle.

The main difference is the size of the anisotropy, which reaches about 1.3 in the
MMS interval compared with about 2 in the simulation. This is partly due to the
strong initial ion anisotropy in the simulation, $T_{i\perp}/T_{i\parallel} = 10$,
which drives ion cyclotron waves with larger amplitudes than in typical
magnetosheath conditions, and thus a stronger compression and trapping of the
electrons. This choice mainly sets the amplitude of the ion-scale structures
rather than the nature of the electron response: the electrons start isotropic,
the ion anisotropy has relaxed to $T_{i\perp}/T_{i\parallel} \approx 2$, within
the range observed in the magnetosheath, before the trapping and the main
whistler activity develop (Section~\ref{sec:evolution}), and the electron
response throughout follows the expected adiabatic, trapping and quasilinear
behaviour (Sections~\ref{sec:evdf}--\ref{sec:whistler_aniso}). In addition, the
observed dips are likely mirror-mode structures, whereas the dips in our
simulation are formed by compressional ion cyclotron waves. Although the
electrons respond to both in a similar way, as argued above, the two types of
structure can differ in depth, shape and evolution, which affects how strongly
the electrons are trapped.

In mirror modes, the trapped particles and the magnetic field evolve together.
\citet{kivelson_mirror_1996} argued that at the field maxima, trapping alone
lowers the pressure enough for marginal stability, whereas in the wells the
trapped particles must also be cooled as their mirror points move apart. We
expect a similar feedback in the compressional ion cyclotron waves studied here.
Test-particle models with a prescribed evolution of the magnetic field can
reproduce butterfly distributions \citep{Li2021}, but cannot include this
feedback, nor the whistlers that the trapped electrons excite, which in our
simulations reshape the distribution.

The simulations use a reduced mass ratio, $m_i/m_e = 100$, which brings ion and
electron scales closer together than in a real plasma and may make the
coupling between them more efficient.  \citet{Ley2024} found
that the secondary whistlers in mirror modes depend only weakly on the mass ratio
between 8 and 64, which suggests that the qualitative picture here should hold,
although the time scales and wave spectra may change.  The 2D geometry also limits which wave
vectors are allowed. 3D simulations with a more realistic mass ratio, however, represent a significantly greater numerical challenge, and are outside the scope of the present work.

\section{Conclusion}
We have used fully kinetic particle-in-cell simulations to study how ion- and
electron-scale instabilities are linked through electron trapping. The only
initial source of free energy is an ion temperature anisotropy. This anisotropy
drives large-amplitude ion cyclotron waves, which become partly compressional and
form magnetic minima and maxima. In turn, these magnetic structures modify the
electron distribution, which subsequently becomes unstable to whistler waves.

At first, the electrons respond adiabatically to the compression, and their anisotropy follows the double-adiabatic prediction. As the structures grow, electrons with small pitch angles escape from the dips along the field, while the trapped electrons remain. The anisotropy is then highest in the dips, and the distributions there develop butterfly shapes similar to those observed by MMS. 
The sharp gradients at the trapping boundary drive whistler waves through cyclotron resonance. The whistlers propagate away from the dips and
scatter the electrons back toward smaller pitch angles, flattening the
distribution along the quasilinear diffusion curves. When the whistler power peaks, the electron anisotropy drops to about half of its adiabatic value, mainly through parallel heating, and settles near the whistler instability threshold.

Our work thus illustrates one pathway through which energy is transported from
ion to electron scales, with the electron velocity distribution as the link. The
shape of the magnetic structure determines the trapping, the trapping determines
the velocity-space gradients, and the gradients determine the whistler emission. As a consequence, bi-Maxwellian instability thresholds alone are not sufficient
to predict whistler generation in magnetic dips; instead the full shape of the electron
distribution must be taken into account.

\section*{Acknowledgments}
The authors are grateful to
Daniel Graham, Jimmy Juno, Yuri Khotyaintsev,
Louis Richard and Andris Vaivads for fruitful discussions. The computations and data handling were enabled by resources provided by the National Academic Infrastructure for Supercomputing in Sweden (NAISS), partially funded by the Swedish Research Council through grant agreement no.~2022-06725. This research was supported by the International Space Science Institute (ISSI) in Bern, through the ISSI International Team projects \#23-588 ('Unveiling Energy Conversion and Dissipation in Non-Equilibrium Space Plasmas') and \#25-653 ('Impact of stellar wind plasma turbulence on planetary magnetospheres in the heliosphere and beyond').

\section*{Funding}
This project received funding from the Knut and Alice Wallenberg Foundation (Grants No. KAW 2022.0087 and KAW 2023.0249) and the Swedish Research Council (Dnrs.~2021-03943 and 2022-02862).

\section*{Declaration of interests}
The authors report no conflict of interest.
\bibliographystyle{jpp}

\bibliography{jpp-instructions}

\end{document}